\documentclass[preprint,preview,12pt]{elsarticle}

\usepackage{graphics}
\usepackage{graphicx}

\usepackage{amssymb}

\usepackage{latexsym}
\usepackage{makeidx}
\usepackage{multirow}
\usepackage{multicol}
\usepackage{float}
\usepackage{subfigure}
\usepackage[ruled, vlined]{algorithm2e}
\usepackage{color, colortbl}
\usepackage{xcolor}

\biboptions{square,numbers, comma, sort&compress}

\begin{document}

\begin{frontmatter}



\title{Report: Performance comparison between C2075 and P100 GPU cards using cosmological correlation functions}


\author[mcm]{Miguel C\'ardenas-Montes}
\address[mcm]{CIEMAT, Department of Fundamental Research. \\Avda. Complutense 40. 28040. Madrid, Spain.}
\ead{miguel.cardenas@ciemat.es}
\author[mcm]{Iv\'an M\'endez-Jim\'enez}
\author[mcm]{Juan Jos\'{e} Rodr\'{i}guez-V\'{a}zquez}
\author[mcm]{Jos\'e Mar\'ia Hern\'andez Calama}

\begin{abstract}
In this report, some cosmological correlation functions are used to evaluate the differential performance between C2075 and P100 GPU cards. In the past, 
the correlation functions used in this work have been widely studied and exploited on some previous GPU architectures. The analysis of the 
performance indicates that a speedup in the range from 13 to 15 is achieved without any additional optimization process for the P100 card.
\end{abstract}


\end{frontmatter}


\section{Introduction}
The increment of the data volume in Cosmology is forcing to implement new solutions for solving the difficulties associated to the large processing time 
and the precision of the calculation. The study of the Large-Scale Structure of the Universe is a discipline where the increment in the data volume has 
been remarkable. In comparison with the previous generation, experiments such as: Dark Energy Survey \citep{DESastro-ph0510346}, Physics of Accelerated 
Universe (PAU) \cite{2009ApJ...691..241B}, Kilo-Degree Survey \cite{KIDSastro-ph1206.1254} and Euclid \cite{laureijs2011,Euclidastro-ph1206.1225} are 
increasing the available data volume for correlation-function calculations a few orders of magnitude.

With this increment, two barriers appear when analysing large data volumes. On the one hand, it is obvious that the increment in the data volume and the 
size of the standard sample produce a correlated increment in the processing time. On the other hand, the increment in the size of the standard sample, 
which has to be analysed as a whole, raises the pressure over the weaknesses of the floating-point representation. 

In order to overcome the barrier of the processing time, two strategies have been followed: the adaptation of the problem to more powerful hardware or 
the use of mathematical artefacts to reduce the computational complexity of the calculation. So, efforts to profit from the computational capacity of 
graphics processing units (GPU) for calculating large data samples have been made. They mainly focus on the two-point angular-correlation function 
\cite{DBLP:conf/caepia/MontesVSPRA13,CardenasMontesIbergrid13,CardenasMontesIbergrid12,2012arXiv1204.6630P,Roeh:2009:ACD:1513895.1513896}, the 
shear-shear correlation function \cite{CardenasMontesCPC14,MontesRSAPRB14}, the aperture mass statistic \cite{Bard:2012cb} and the two-point 
three-dimensional correlation function \cite{Mendez-Jimenez:2017:AIT:3101112.3101253}. 

Additionally to the larger processing time, the increment in the size of the standard sample carries out precision weaknesses associated with the number 
representation. 
The size of the standard sample forces to use parallel processing for achieving acceptable times. Most of the success cases corresponds to 
GPU computing. However, it has been demonstrated how the traditional histogram-construction algorithm on GPU \cite{Sanders2010CUDA}, when processing 
large samples, introduces errors in the most populated bins. For overcoming this barrier, alternative implementations for histogram construction on 
GPU ---termed \textit{bin-recycling strategy}--- have been proposed \cite{CardenasMontesRV16}. An example of the implementation of the \textit{bin-recycling
strategy} in the 2PACF can be found in \cite{DBLP:conf/ica3pp/MontesRVSG16}.

The final aim for this test is to evaluate the speedup of the P100 GPU card in relation with the previous works performed on the C2075 card. 

The rest of the document is organized as follows: Section \ref{section:relatedwork} summarizes related work and previous efforts made in this area.  
An overview of the estimators of the Large-Scale Structure of the Universe and their computational complexity is presented in Sections \ref{section:MM:2PACF}, 
\ref{section:MM:2P3DCF}, \ref{section:MM:3PACF}. 
The commonalities of the histogram construction on GPU are described in Section \ref{section:MM:commonalities}. 
The weaknesses of floating-point representation are described in Section \ref{section:MM:weaknesses}. 
The Bin-Recycling Strategy is briefly explained in Section \ref{section:MM:binrecycling}.
The Results and the Analysis are shown in Section \ref{section:results}. Finally, the Conclusions are presented in Section \ref{section:conclusions}.

\section{Related Work}\label{section:relatedwork}
Previous efforts made to reduce the processing times of correlation functions have focused on two strategies. On the one hand, the use of space division 
algorithms  to avoid calculating all the possible combinations of $m$ galaxies (multiplets of galaxies). In this case, two algorithms have been proposed 
to reduce the computational complexity of the correlation functions: kd-tree \cite{Moore_Connolly_Genovese_Gray_Grone_Kanidoris_Nichol_Schneider_Szalay_Szapudi_etal._2000} 
and pixelization \cite{0067-0049-151-1-1}. 

With the pixelization technique, the sky is divided following a fixed pixelization scheme. This division does not take into account where the galaxies are and 
one cell is contiguous to the others. By dividing the space in pixels, this technique allows to reduce the processing time from $O(N^m)$ to $O(N_{pix}^m)$. 
Since the number of pixels is much lower than the number of galaxies, this largely reduces the processing time. 

In the kd-tree approach, the galaxies are organized in a hierarchy of bounding boxes. This structure is used for discarding the galaxies which the 
closest corners of the boxes sustain an angle larger than the largest angle of the histogram.

Two uses of space-division algorithms are possible. Once the space has been divided in cells, the variable under study is averaged for each cell, so this mean 
value becomes the cell value. One of the uses of these algorithms is to calculate the galaxies individually below a threshold and the average value in the cell 
above the threshold. This threshold can be a distance or an angle. Oppositely, in the second case the threshold is used to discard bulks of calculations. As a result, 
a reduction of the processing time is achieved in both.

On the other hand, the following algorithms also benefitted from the performance GPUs can attain: the 2PACF \cite{DBLP:conf/caepia/MontesVSPRA13,CardenasMontesIbergrid13,CardenasMontesIbergrid12,2012arXiv1204.6630P,Roeh:2009:ACD:1513895.1513896}, the  shear-shear 
correlation function \cite{CardenasMontesCPC14,MontesRSAPRB14}, the aperture mass statistic \cite{Bard:2012cb} and the two-point three-dimensional correlation 
function \cite{Mendez-Jimenez:2017:AIT:3101112.3101253}. In these cases, the throughput-oriented processor architecture of the GPUs is used for exploiting the abundant 
parallelism of the correlation functions. 

Other aspect that should be underlined is the precision in the histogram calculation. When analysing large input files, issues related to the largest 
representable number and the largest consecutively-representable integer become relevant. In the past, strategies to modify the standard algorithm 
\cite{Sanders2010CUDA} for histogram construction in the cosmological correlation-function context have been proposed \cite{CardenasMontesRVSA14}. 

Due to the limitation of energy consumption and space in the computational facilities of the research institutions, studies about the energy to solution 
have also been undertaken using the cosmological correlation functions \cite{ASHeS2017}. 

Concerning the floating-point format and its weaknesses, a complete description can be found in \cite{Goldberg:1991:CSK:103162.103163}. In 
\cite{Whitehead:FitFlorea:PrecisionGPU}, the inexactnesses associated with the use of float representation (operation accuracy and rounding) on GPU, 
and how the programming affects the final result, is presented. Finally, in \cite{4610935} the latest version of the floating-point standard can be found.

\section{Methods and Materials\label{section:MM}}

\subsection{Two-Point Angular-Correlation Function\label{section:MM:2PACF}}
The two-point angular-correlation function (2PACF), $\omega(\theta)$, is a computationally intensive function which measures the excess or lack of 
probability of finding a pair of galaxies separated by a certain angle $\theta$ with respect to a random distribution. This function has a computational 
complexity of $O(N^2)$, where $N$ is the number of galaxies. Diverse estimators exist for 2PACF, the estimator proposed by \cite{1993ApJ...412...64L}, 
(Eq. \ref{eq:2PACF}), is the most widely used by cosmologists by virtue of its minimum variance.

\begin{equation}
\omega(\theta) = 1 \;+\; \left(\frac{N_{random}}{N_{real}}\right)^2 \cdot \frac{DD(\theta)}{RR(\theta)} \;-\; 2 \cdot \frac{N_{random}}{N_{real}} \cdot \frac{DR(\theta)}{RR(\theta)}
\label{eq:2PACF}
\end{equation}
where

\begin{itemize}
\item DD($\theta$) is the number of pairs of galaxies for a given angle $\theta$ chosen from the observational data catalogue, D, with $N_{real}$ galaxies.
\item RR($\theta$) is the number of pairs of galaxies for a given angle $\theta$ chosen from the random catalogue, R, with $N_{random}$ galaxies.
\item DR($\theta$) is the number of pairs of galaxies for a given angle $\theta$ taking one galaxy from the observational data catalogue D and another 
from the random catalogue R.
\end{itemize}


Beyond the study of the Large Scale Structure of Universe, the correlation functions can be employed in ecology for species correlation and text mining. 
Therefore, the improvement developed in this work can be applied to these analyses.

\subsection{Two-Point Three-Dimensional Angular-Correlation Function\label{section:MM:2P3DCF}}
The 2P3DCF, $\omega(\theta,\delta z)$, evaluates the excess or lack of probability of finding a pair of galaxies separated by a certain angle $\theta$ and 
a certain redshift difference $\delta z$ with respect to a random distribution. There is a relationship between the redshift of a galaxy and its distance 
to the observer, and therefore in this work the redshift difference should be understood as a measure of the distance between two galaxies along the line 
of sight. Equally to 2PACF, it has a computational complexity of $O(N^2)$ where $N$ is the number of galaxies.

\begin{equation}
\omega(\theta,\delta z) = 1 \;+\; \left(\frac{N_{random}}{N_{real}}\right)^2 \cdot \frac{DD(\theta,\delta z)}{RR(\theta,\delta z)} \\
\;-\; 2 \cdot \frac{N_{random}}{N_{real}} \cdot \frac{DR(\theta,\delta z)}{RR(\theta,\delta z)}
\label{eq:2P3DCF}
\end{equation}
where:
\begin{itemize}
\item $DD(\theta,\delta z)$ is the number of pairs of galaxies with a separation $(\theta,\delta z)$ where the pairs of galaxies have been chosen from 
the observational catalogue. This catalogue includes $N_{real}$ galaxies.
\item $RR(\theta,\delta z)$ is equivalent to $DD(\theta,\delta z)$ but selecting the galaxies from the random catalogue. This catalogue includes $N_{random}$ 
galaxies.
\item $DR(\theta,\delta z)$ is equivalent to the two previous values but forming the pairs with one galaxy from the real catalogue and the other one from 
the random catalogue. 
\end{itemize}

\subsection{Three-Point Angular-Correlation Function\label{section:MM:3PACF}}
The three-point angular-correlation function (3PACF), $\zeta(\theta)$, is similar to the 2PACF but involving three galaxies instead of two. This modification 
increases the computational complexity to $O(N^3)$, where $N$ is the number of galaxies. In this case, one of the most used estimators is the one proposed by 
\cite{1998ApJ...494...41L} (Eq. \ref{eq:3PCF}).

The three-point angular-correlation function (3PACF),$\zeta(\theta)$, measures the excess or lack of probability of finding a triplet of galaxies separated by 
a certain triplet of angles $\theta_1\theta_2\theta_3$ with respect to a random distribution.

\begin{equation}
\zeta(\theta) = \left(\frac{N_{random}}{N_{real}}\right)^3 \cdot\frac{DDD}{RRR} \,-\, 3 \cdot \left(\frac{N_{random}}{N_{real}}\right)^2 \cdot\frac{DDR}{RRR} \,+\, 3 \cdot \frac{N_{random}}{N_{real}} \cdot\frac{DRR}{RRR} \,-\, 1
\label{eq:3PCF}
\end{equation}
where

\begin{itemize}
\item $DDD(\theta_1\theta_2\theta_3)$ denotes the number of triplets of galaxies for a given set of angles $\theta_1\theta_2\theta_3$ where the three galaxies 
are selected from the observational data catalogue D.
\item $RRR(\theta_1\theta_2\theta_3)$ denotes the number of triplets of galaxies for a given set of angles $\theta_1\theta_2\theta_3$ where the three galaxies 
are selected from the random data catalogue R.
\item $DDR(\theta_1\theta_2\theta_3)$ and $DRR(\theta_1\theta_2\theta_3)$ are similar to the previous ones taking two galaxies from one catalogue and the third 
one from the other catalogue.
\item $N_{real}$ and $N_{random}$ are the number of galaxies in the observational (D) and random (R) data catalogues respectively.
\end{itemize}

The three-point angular-correlation function contains information about the initial conditions of the Universe, in the sense that a non-vanishing value of this 
quantity can point to non-Gaussian distribution of the initial perturbations, which is what is predicted by the most simple 'inflation' models we 
currently use as a basis for our understanding of the early development of the Universe. If this picture can not be described by Gaussian perturbations, 
cosmology would have to make a major overhaul in understanding the first instants of the Universe. In addition, 2PACF information is degenerate with the 
so-called "galaxy bias": a constant offset between the matter and galaxy clustering which is removed by considering the additional information provided by 
3PACF \cite{Szapudi2009}. On smaller scales, 3PACF also has a link with galaxy formation and evolution \cite{1993ApJ...413..447F}. 

An in-depth explanation about the correlation functions used in this work and their relevance in the cosmological studies for analysing the Large Scale Structure 
of Universe can be found in the bibliography \cite{Fu-2008,2013MNRAS...430...2200}.

\subsection{Commonalities in the Histogram Construction on GPU\label{section:MM:commonalities}}
The implementation proposed in \cite{Sanders2010CUDA} is accepted as the most conventional for histogram construction on GPU (Algorithm \ref{alg:pseudocode:basic}). 
Among its positive features, this implementation holds a great flexibility to be adapted to different types of problems along with a low complexity which implies that it 
is easily understandable and adaptable.

\begin{algorithm}[!ht]
\ForEach{Pair of Points}{
	Calculate the coincidence function\;
	\If{Coincidence value is in the histogram's range}{
	Add the count value to the appropriate bin of the sub-histogram on \textit{shared memory}. Atomic addition is required\;
}
}
$imax=\frac{histogram size}{threads per block}$\;
\While{i$\le$imax}{
	Accumulate the sub-histograms in the single final histogram on \textit{global memory}. Atomic addition is required\;

}
\caption{Algorithm pseudocode for the shared-memory-based implementation for building histograms on GPU without the \textit{bin-recycling strategy}\label{alg:pseudocode:basic}}
\end{algorithm}

The kernel flow of this implementation is as follows. At a given time, each thread is acting on a set of objects. For each pair of objects, the coincidence 
function is calculated. If the coincidence value is in the histogram's range, then a count is added to the appropriate bin. For 2PACF, the count value is the 
unit and the coincidence function is the angular distance between each pair of galaxies.

Due to the parallel nature of the GPU architecture, race conditions can be generated. Several threads in a thread block might try to update the number of counts 
in the same bin. Therefore, an atomic operation for the addition of the count value is necessary; concretely, the \texttt{atomicAdd()} function is required.

In order to increase the degree of parallelism, each thread block has its sub-histogram allocated on shared memory. Threads of different blocks act on 
the number of counts of their own sub-histogram in parallel. If two threads of a block update the number of counts of two bins, the atomic function performs 
the operation in parallel; whereas updates in the same bin are executed sequentially. 

Once all the coincidences have been calculated and the counts stored in the sub-histograms, they are accumulated in the final histogram on global memory. For 
this addition, the use of \texttt{atomicAdd()} function is also proposed.

Presently, the \texttt{atomicAdd()} function is available for the following 32-bit types: integer, unsigned integer and floating point; and 64-bit types: unsigned-long-long integer. The absence of an implementation of 64 bits for floating-point representation and the large processing time when using 64-bit integer \texttt{atomicAdd()} function have motivated the search of alternative methods for processing large samples \cite{CardenasMontesCPC14,CardenasMontesRV16}. 

This schema for calculating histograms on GPU can be easily translated ---modifying the mechanism of the coincidence calculation and the coincidence value--- to other 
disciplines where information is presented as histograms.

\subsection{Weaknesses of the Floating-Point Representation\label{section:MM:weaknesses}}
By using floating-point arithmetic, some limitation on the representation of the figures is assumed. Both single ---32 bits--- and double ---64 bits--- floating-point representation is limited in the smallest and the largest representable number. Contrary to the integer case, in the floating-point representation not all the integers in the range between the minimum and the maximum number are representable, regardless the number of bits in the representation. 

Since the count value is the unit in 2PACF, the largest-con\-secu\-tively-repre\-sentable integer in floating point becomes relevant for the final 
number of counts accumulated in the most populated bins. In 32-bit floating point, the largest-con\-secu\-tively-repre\-sentable integer is 16777216, whereas in 
64 bits is 9007199254740992. Considering the kernel flow of the histogram construction on GPU (Section \ref{section:MM:commonalities}), it is seen that as far as 
the bins are populated and the number of accumulated counts approaching to the largest-con\-secu\-tively-repre\-sentable integer in floating point, the risk of 
coincidence loss becomes critical. Once a bin has reached this figure, any additional coincidence does not increase the value of counts in the bin. Therefore, the 
most populated bins are limited in the accumulated counts. A priory, these highly populated bins can not be predicted, thus the results are not reliable. 

This risk implies that once a bin has reached the largest-con\-secu\-tively-repre\-sentable integer, the addition of a count in this bin does not produce any 
increment. Since the most populated bins can not be predicted in most of the cases, the results are not reliable. 

In 2PACF, the kernel (Algorithm \ref{alg:pseudocode:basic}) goes through all the pairs of galaxies by using a double loop ---while-for. Each thread of a block 
composes the pairs of galaxies executing a for-loop. So, all the threads of a block accumulate the coincidences in their own sub-histogram in shared memory. In the 
most critical scenario where all the coincidences feed a single bin, the number of counts can be calculated \footnote{The analysis of the implementation for files 
containing all the objects in a single position is used in this part of the work  \cite{CardenasMontesCPC14,CardenasMontesRV16}. This case is the most critical configuration 
since it largely increases the stress over the floating-point representation. At the same time, it allows to verify the correctness of the final results.}. In our 
studies \cite{CardenasMontesCPC14,CardenasMontesRV16} this issue is illustrated as follows, a configuration of 64 threads per block is used. If the standard sample size is $10^6$ galaxies, then these threads generate $64\times 10^6$ coincidences. 
This result overflows the largest-consecutively-representable integer in 32-bit floating-point representation. 

In order to overcome this overflow, unsigned-long-long-integer \texttt{atomicAdd()} can be used. However, it has been demonstrated that the use of this representation 
critically impacts the processing time \cite{CardenasMontesRV16}.

A secondary aspect of floating-point representation is the lack of some of the integers in the range 16777216 to the maximum representable number, 
$\approx 3.4 \cdot 10^{38}$. In 32-bit floating-point representation, from $2^{24}$ to $2^{25}$ every second integer is exactly represented, from $2^{25}$ to $2^{26}$ 
every fourth integer is exactly represented, from $2^{26}$ to $2^{27}$ every eighth integer, and so on \cite{scherer2010computational}.  Due to this, when 
accumulating the sub-histograms in the final one, the sum of the counts might result in a non-representable integer. In this case, the number of counts 
will be rounded to the closest representable integer.

This problem has been partially mitigated in \cite{CardenasMontesRVSA14} and fully mitigated in \cite{CardenasMontesRV16} (see Section \ref{section:MM:binrecycling}).

\subsection{Bin-Recycling Strategy\label{section:MM:binrecycling}}
As was mentioned, the construction of histograms on GPU encompasses two phases: the calculation of the coincidences with the population of the sub-histograms and the 
accumulation of the sub-histograms in the final one (Algorithm \ref{alg:pseudocode:basic}). 

In the case where each coincidence is signalised with a unit value and a very large sample size is analysed, the number of coincidences can overflow the 
largest-con\-secu\-tively-repre\-sentable  integer. In order to avoid this overflow, the \textit{bin-recycling strategy} proposes including the accumulation of 
the sub-histograms in the calculation of the coincidences without waiting for the end of this phase \cite{CardenasMontesRV16}. The application of the \textit{bin-recycling 
strategy} to the 2PACF divides the calculation of the coincidence phase in some parts, accumulating the count store in an array of sub-histograms on \textit{global memory}, 
and then resetting to zero the bins of the sub-histograms (Algorithm \ref{alg:pseudocode:merged}). Later, the sub-histograms are accumulated in the final histogram on 
the host's memory. 
With the appropriate implementation, this mechanism can avoid reaching the limit of the largest-con\-secu\-tively-repre\-sentable integer (16777216 for single-precision 
floating point).

\begin{algorithm}[!ht]
\ForEach{Pair of Points}{
	Calculate the coincidence function\;
	\If{Coincidence value is in the histogram's range}{
	Add the count value to the appropriate bin of the sub-histogram on shared
	memory. Atomic addition is required\;
}
$imax=\frac{histogram size}{threads per block}$\;
\While{i$\le$imax}{
	Copy sub-histograms on \textit{global memory}\;
        Bin-recycling mechanism (all the bins are restarted to zero)\;
}
}
Accumulate the sub-histograms in the final histogram on the host's memory\;
\caption{Algorithm pseudocode for the shared-memory-based implementation for building histograms on GPU with the \textit{bin-recycling strategy}\label{alg:pseudocode:merged}}
\end{algorithm}

In correlation functions of higher order, such as the Three-Point Angular-Correlation Function (3PACF), the multiple loops permit to place the \textit{bin-recycling 
strategy} in diverse positions in relation to the loops. However, to activate this possibility in the 2PACF the for-loop has to be divided in several chunks, and the 
\textit{bin-recycling mechanism} be placed at the end of these chunks. An in-depth analysis for the application of the \textit{bin-recycling strategy} to the 3PACF 
can be found in \cite{CardenasMontesRV16}.

\section{Results and Discussion\label{section:results}}

All the numerical experiments have been executed on the following two machines: 

\begin{itemize}
\item Dual Intel Xeon X5570 at 2.93 GHz, 8 GB of RAM and one Nvidia C2075 GPU card with Fermi architecture.
\item Dual Intel Xeon E5-2650v4 at 2.20 GHz, 128 GB of RAM and four Nvidia P100 SXM2 GPU cards with Pascal architecture (just one card was used).
\end{itemize}

Next, the processing time of the most stringent cases of the 2PACF, 2P3DCF and 3PACF from the literature are compared with the processing time on the P100 card.

\subsection{Execution Time of the Two-Point Angular-Correlation Function\label{section:2PHAccACF}}
In \cite{DBLP:conf/ica3pp/MontesRVSG16}, an implementation of the 2PACF using the \textit{bin-recycling strategy} \cite{CardenasMontesRV16} is presented. 
The \textit{Bin-recycling strategy} inherits the efforts for producing accuracy-aware implementations \cite{CardenasMontesRV16,CardenasMontesRVSA14}. In this section, the 
most relevant results in \cite{DBLP:conf/ica3pp/MontesRVSG16}, carried out on a C2075 GPU card, are compared with a P100 GPU card (Table \ref{table:2PHAccACF:ET}).

\begin{table*}
\caption{Execution-time (ms) comparison between the C2075 and P100 cards for the 2PACF with the \textit{bin-recycling strategy} using one and five million galaxies 
 respectively.}
\label{table:2PHAccACF:ET}
\centering
\begin{tabular}{rrrr}\hline
Input Size & C2075 & P100 & Speedup \\ \hline
$10^6$ & $1224893\pm1136$ & $88980\pm277$ & $13.8$ \\ 
$5 \cdot 10^6$ & -- & $2202438\pm1598$ & -- \\\hline
\end{tabular}
\end{table*}

When processing one million galaxies from CFHTLenS on the P100 (Table \ref{table:2PHAccACF:ET}), a speed-up of $13.8$ is achieved with respect to the C2075\footnote{In our study, 
a configuration of 64 threads per block and 32-bit floating-point representation are used.}. This processing time is obtained with no code optimisation for the newer 
architecture.

The gain obtained in the P100 invites to test larger input files. For this purpose, a five-times-copy file is created using an input size of one million galaxies as a base. For this five-million-galaxy input the 2PACF takes $2202438\pm1598$. 

These last results show that analysing input sizes in the order of the tens of millions of galaxies is feasible with a P100 GPU card; and potentially, a hundred million galaxies 
for a multiple card system. These input sizes are expected in the near future for cosmological catalogues.

\subsection{Execution Time of the Two-Point Three-Dimensional Correlation Function\label{section:2P3DCF}}
An implementation of the 2P3DCF was presented in \cite{Mendez-Jimenez:2017:AIT:3101112.3101253}. In this section, the most relevant test of that 
previous work, which were run back then on a C2075 card, has been re-evaluated on a P100 card (Table \ref{table:2P3DCF:ET}). The experiments regarding this section have been 
executed for an input file with 1086470 galaxies using 64 bins per angle, one bin per redshift and 64 threads per block.

\begin{table*}
\caption{Execution-time (ms) comparison between C2075 and P100 cards for the 2P3DCF using an input size of 1086470 galaxies.}
\label{table:2P3DCF:ET}
\centering
\begin{tabular}{rrr}\hline
 C2075 & P100 & Speedup\\\hline
 $1521139\pm584$ & $101136\pm198$ & $15.04$ \\ \hline
\end{tabular}
\end{table*}

The comparison between the processing times on the C2075 and P100 cards achieves a speedup of $15.04$, which is similar to the figure of the 2PACF (Table 
\ref{table:2PHAccACF:ET}).

\subsection{Execution Time of the Three-Point Angular-Correlation Function\label{section:3PACF}}
In comparison with the correlation function involving two galaxies, the 3PACF has a larger complexity and, therefore, a much larger execution time is obtained even for humble 
input sizes. Furthermore, the larger number of coincidences to store in the histograms for an equal input size puts some additional pressure on the accuracy 
of the calculation. In \cite{CardenasMontesRVSA14,CardenasMontesRV16}, efforts to maintain the accuracy in the 3PACF case are shown. These studies 
focus on the calculation of a single histogram from the Eq. \ref{eq:3PCF}, e.g. DDD, instead of the whole estimator. Until this moment, the full calculation of 
the 3PACF for one million galaxies had not been tackled.

In this section, the largest-input cases in \cite{CardenasMontesRV16}, which were executed on the C2075 card, are compared with the execution times on the P100 card (Table 
\ref{table:3PACF:DDD:ET}). 

\begin{table*}
\caption{Execution Times (ms) for the internal placement of the bin-recycling mechanism and the schema \textit{while-while-for} for diverse grids of blocks. The largest 
configuration for each input size corresponds to the case where the limit of active thread blocks is pointed by $\lceil{\frac{input\,size}{threads\,per\,block}\rceil}$.}
\label{table:3PACF:DDD:ET}
\centering
\begin{tabular}{r rr rr} \hline
 \multirow{3}{*}{Points} & \multicolumn{2}{c}{Execution time} & Speedup \\
 & C2075 & P100 &&  \\ \hline
 $4\cdot10^4$ & 47036124 & 3345782 & 14.06  \\ 
 $10^5$ &  & 52313916 & \\\hline
\end{tabular}
\end{table*}


When comparing the processing time for an input size of $4\cdot10^4$ galaxies, the speedup obtained by the P100 in relation with the C2075 card is 14.06 if the galaxies are 
randomly distributed in the analysed sky area. This value is similar to those obtained for the 2PACF and one-million-galaxy input. 

In \cite{CardenasMontesRV16}, comparisons with an MPI implementation of the 3PACF are also performed\footnote{For this comparison, an MPI implementation using 64-bit 
representation has been implemented and executed in a cluster composed by 144 nodes with two quad-core Xeon processors at 3.0 GHz with 8 GB of RAM at 667 MHz.}. In Table 
\ref{table:3PACF:DDD:MPIComparison}, the processing times for diverse number of cores in the MPI implementation as well as those corresponding to the execution on 
the C2075 and P100 cards are shown.

\begin{table*}
\caption{Execution-time (s) comparison between GPU (C2075 and P100) and MPI implementations for the DDD calculation of the 3PACF and $10^4$ galaxies as input size. Values 
are presented in descending processing time.}
\label{table:3PACF:DDD:MPIComparison}
\centering
\begin{tabular}{lr}\hline
Hardware& Execution Time \\  \hline
MPI-64  & 959 \\  
C2075   & 731 \\
MPI-128 & 485 \\ 
P100    & 124 \\ 
MPI-256 &  77 \\ 
\hline
\end{tabular}
\end{table*}

Comparing with the 64-bit MPI implementation of the 3PACF (Table \ref{table:3PACF:DDD:MPIComparison}), it is appreciated that the MPI implementation outperforms the 
C2075 bin-recycling strategy when using at least 128 cores; whereas for fewer cores, the GPU implementation takes shorter than the MPI one. Otherwise, if the 
comparison is made with the P100 implementation, then this outperforms also the MPI implementation with 128 cores\footnote{Due to restrictions in the 
use of the cluster facility, executions using a larger number of cores are not allowed.}.

\section{Conclusions}\label{section:conclusions}
In this work, a review of the relative performance between GPU cards: C2075 and P100, for diverse cosmological correlation functions has been presented. Such functions 
have a high computational intensity and for this reason they have been successfully implemented on GPU cards.

The processing-time comparison between the C2075 and P100 cards indicates a speedup in the range from 13 to 15. Taking into account that the codes have
been compiled without any additional optimizations, the obtained speedup is remarkable.

\section*{Acknowledgements}

The research leading to these results has received funding by the Spanish Ministry of Economy and Competitiveness (MINECO) through the grants FPA2013-47804-C2-1-R, 
FPA2016-80994-C2-1-R and MDM-2015-0509 ("Unidad de Excelencia Mar\'ia de Maez\-tu": CIEMAT - F\'ISICA DE PART\'ICULAS).

IMJ is co-funded in a 91.89 percent by the European Social Fund within the Youth Employment Operating Program, for the programming period 2014--2020, as well as Youth Employment Initiative (IEJ). IMJ is also co-funded through the Grants for the Promotion of Youth Employment and Implantation of Youth Guarantee in Research and Development and Innovation (I+D+i) from the MINECO.

The CFHTLens data are based on observations obtained with the Mega-Prime/MegaCam, a joint project of CFHT and CEA/DAPNIA, at the Canada-France-Hawaii Telescope (CFHT) which is operated by the National Research Council (NRC) of Canada, the Institut National des Sciences de l'Univers of the Centre National de la Recherche Scientifique (CNRS) of France and the University of Hawaii. This research used the facilities of the Canadian Astronomy Data Centre operated by the National Research Council of Canada with the support of the Canadian Space Agency. The CFHTLenS data processing was made possible thanks to significant computing support from the NSERC Research Tools and Instruments grant program.

\end{document}